\documentclass[11pt]{article}
\usepackage{amsmath,amssymb,color,graphics,epsfig}
\usepackage{amsfonts}
\newcommand{\be}{\begin{equation}}
\newcommand{\ee}{\end{equation}}
\newcommand{\bea}{\setlength\arraycolsep{2pt} \begin{eqnarray}}
\newcommand{\eea}{\end{eqnarray}}
\newcommand{\nn}{\nonumber}

\def\ft#1#2{{\textstyle{\frac{\scriptstyle #1}{\scriptstyle #2} } }}

\def\0{{\sst{(0)}}}
\def\1{{\sst{(1)}}}
\def\2{{\sst{(2)}}}
\def\3{{\sst{(3)}}}
\def\4{{\sst{(4)}}}
\def\5{{\sst{(5)}}}
\def\6{{\sst{(6)}}}
\def\7{{\sst{(7)}}}
\def\8{{\sst{(8)}}}
\def\sst#1{{\scriptscriptstyle #1}}

\begin{document}

\begin{flushright}
%\hfill{KIAS-P12028}
 %\hfill{
%\bf hep-th/yymmnnn}
\end{flushright}

\vspace{25pt}
\begin{center}
{\large {\bf Non-Abelian (Hyperscaling Violating) Lifshitz Black Holes\\ in General Dimensions \\ }}

\vspace{10pt}
Xing-Hui Feng and Wei-Jian Geng

\vspace{10pt}

{\it Department of Physics, Beijing Normal University, Beijing 100875, China}

\vspace{40pt}

\underline{ABSTRACT}
\end{center}

We consider Einstein gravities coupled to a cosmological constant and multiple $SU(2)$ Yang-Mills fields in general dimensions and find that the theories admit colored Lifshitz solutions with dynamic exponents $z>1$.  We also introduce a Maxwell field and construct exact electric charged black holes that asymptote to the $z=D-1$ colored Lifshitz spacetimes and analyse their thermodynamical first law. Furthermore, we introduce a dilaton to the system and construct Lifshitz spacetimes with hyperscaling violations. After turning on the Maxwell field, we obtain a class of hyperscaling violating Lifshitz black holes when $\theta=\frac{2}{D-2}[z-(D-1)]$.

\vfill {\footnotesize Emails: xhfengp@gmail.com \ \ \ vergilgeng@gmail.com }

\thispagestyle{empty}

\pagebreak
%\voffset=0pt
%\setcounter{page}{1}

%\tableofcontents
%\addtocontents{toc}{\protect\setcounter{tocdepth}{2}}

%%%%%%%%%%%%%%%%%%%%%%%%%%%%%%%%%%%%%%%%

\newpage
%%%%%%%%%%%%%%%%%%%%%%%%%%%%%%%%%%%%%%%%

\section{Introduction}

Holographic technique provides a powerful and wide-ranging tool to study strongly-coupled field theories by embedding them in the boundaries of some gravitational backgrounds such as the anti-de Sitter (AdS) spacetimes. In condensed matter physics the gauge/gravity duality has been applied successfully to - among numerous other systems - quantum critical points exhibiting Lifshitz \cite{Kachru:2008yh} and Schr\"odinger \cite{sd} symmetry. Lifshitz quantum critical points are invariant under the
scaling symmetry
\be
t\rightarrow\lambda^zt,\qquad x^i\rightarrow\lambda x^i, \label{scaling}
\ee
where $z$ is the dynamical critical exponent. In \cite{Kachru:2008yh}, a candidate gravity dual for Lifshitz fixed points was proposed, with metric
\be
ds^2=\ell^2\left(-r^{2z}dt^2+\frac{dr^2}{r^2}+r^2dx^idx^i\right).
\ee
This is the so-called Lifshitz spacetimes. It is invariance under (\ref{scaling}) provided if one scales $r\rightarrow r/\lambda$, where $r$ is the coordinates of the extra dimension. When $z=1$, the metric reduces to the usual AdS metric in Poincar$\acute{\text{e}}$ coordinates with AdS radius $\ell$.

Recently, non-relativistic backgrounds with scale symmetry as a conformal isometry - but not an isometry - were also
put forward as holographic duals to quantum systems exhibiting hyperscaling violation \cite{Charmousis:2010zz,Gouteraux:2011ce,Iizuka:2011hg,Huijse:2011ef}. The hyperscaling violating Lifshitz metric is of the form
\be
ds^2=r^\theta\left(-r^{2z}dt^2+\frac{dr^2}{r^2}+r^2dx^idx^i\right),
\ee
where $\theta$ is the hyperscaling parameter. When $\theta=0$, it reduces to the Lifshitz spacetime.

Lifshitz  and hyperscaling violating Lifshitz backgrounds have been discussed in a number of recent papers, including [7-21, 22-35] and references therein. Especially, some exact charged black holes in $SU(2)$-colored Lifshitz spacetimes were obtained in four and five dimensions \cite{Fan:2014ixa,Fan:hv}. In this theory, Lifshitz spacetimes themselves can be colored in Einstein-Yang-Mills gravity.

The main work of this paper is to generalize the program in \cite{Fan:2014ixa,Fan:hv}. In these two papers, the authors considered Einstein gravity with a cosmological constant minimally coupled to one $SU(2)$ Yang-Mills field. The self-interaction of the Yang-Mills field allows one to find a class of colored Lifshitz vacua.  In this construction, the dimension of the Euclidean space $x^i$ is related to the number of $SU(2)$ generators. Since the group $SU(2)$ has three generators,  it turns out that there are two possible non-trivial choices of the gauge potentials, giving rise to four and five spacetime dimensions \cite{Fan:2014ixa,Fan:hv}. In order to consider non-abelian Lifshitz spacetimes in higher dimensions, we need to consider larger groups.
The simplest way is to generalize the single $SU(2)$ Yang-Mills field to multiple $SU(2)$ fields.  We shall construct such theories in  section 2. Not so surprisingly, we find that the theories admit colored Lifshitz vacua with the scaling exponents $z>1$ in higher dimensions.  Furthermore, we find that introducing a Maxwell field allows us to construct some exact charged black holes in the colored Lifshitz backgrounds.  In section 3, we introduce a dilaton and construct Lifshitz spacetimes with hyperscaling violation.  We conclude the paper in section 4.

\section{Lifshitz black holes in Einstein-Yang-Mills-Maxwell theory}

\subsection{The set up}

We start with Einstein gravity in general dimensions coupled to a cosmological constant and $N$ $SU(2)$ Yang-Mills fields $A^a_I$ $( a=1,2,3$ and $I=1,2,\cdots,N)$ and Maxwell field $\mathcal{A}$. The Lagrangian is
\be
\mathcal{L}_D=\sqrt{-g}\left(R-2\Lambda-\sum^N_{I=1}\frac{1}{2g^2_I}F^2_I-\frac{1}{2}\mathcal{F}^2\right),
\ee
where $F^2=F^a_{\mu\nu}F^{a\mu\nu}$. The Yang-Mills and Maxwell field strengths are defined as
\be
F^a_{\mu\nu}=\partial_\mu A^a_\nu-\partial_\nu A^a_\mu+\epsilon^{abc}A^b_\mu A^c_\nu,\quad \mathcal{F}=d\mathcal{A}.
\ee
The full set of covariant equations of motion are:
\be
\nabla_\mu F^{a\mu\nu}+\epsilon^{abc}A^b_\mu F^{c\mu\nu}=0,\quad \nabla_\mu\mathcal{F}^{\mu\nu}=0,\label{eom21}
\ee
\begin{align}
R_{\mu\nu}=&\frac{2\Lambda}{D-2}g_{\mu\nu}+\sum^N_{I=1}\frac{1}{g^2_I}\left(g^{\mu\rho}F^a_{I\mu\rho}F^a_{I\nu\sigma}-\frac{1}{2(D-2)}F^2_Ig_{\mu\nu}\right)\nn\\
&+\left(g^{\rho\sigma}\mathcal{F}_{\mu\rho}\mathcal{F}_{\nu\sigma}-\frac{1}{2(D-2)}\mathcal{F}^2g_{\mu\nu}\right).\label{eom22}
\end{align}
Note that we have suppressed the index $I$ in the equations of motion for Yang-Mills fields for simplicity, because they take the same form for all Yang-Mills fields.

Now we consider a class of solutions whose metrics take the form
\be
ds^2=-h(r)dt^2+\frac{dr^2}{f(r)}+r^2dx^i_Idx^i_I.
\ee
where $i$ takes values $1,2$ for $I=1,2,\cdots,m$, else $1,2,3$ for $I=m+1,m+2,\cdots, N$. The motivation for putting the indices in this form will be clear presently.

In this paper, we construct charged black holes in Lifshitz vacua in general dimensions.  It is well-known that a Maxwell field with Lifshitz symmetry cannot satisfy its equations of motion. It was shown in \cite{Fan:2014ixa} that Lifshitz vacua can be supported by Yang-Mills fields, owing to their self-interaction.  For general dimensions, we consider multiple $SU(2)$ Yang-Mills fields.  The ansatz for $N$ $SU(2)$ fields is given by
\be
A_I=\begin{cases}
\sum^2_{a=1}\tau^aA^a_I=\sqrt{2}\psi(r)(\tau^1dx^1_I+\tau^2dx^2_I),\quad&I=1,2,\cdots,m\\
\sum^3_{a=1}\tau^aA^a_I=\psi(r)(\tau^1dx^1_I+\tau^2dx^2_I+\tau^3dx^3_I),\quad&I=m+1,m+2,\cdots,N
\end{cases}
\ee
where $\tau^a$ are the Pauli matrices. In order to find certain exact solutions, we find that the coupling constant must satisfy the following relation
\be
g^2_I=\begin{cases}
2g^2_s,\quad&I=1,2,\cdots,m\\
g^2_s,\quad&I=m+1,m+2,\cdots,N
\end{cases}
\ee
To understand this ansatz, we note that turning on a single gauge potential of an $SU(2)$ triplet effectively reduces to a Maxwell field and hence it is ruled out by the equations of motion.  We thus need to turn on two or all three gauge potentials of each $SU(2)$ to make use of the self interaction. This implies that the spacetime dimension is given by
\be
D-2=2m+3n.\label{re1}
\ee
The total number of $SU(2)$ Yang-Mills fields is
\be
N=m+n.\label{re2}
\ee
For low-lying dimensions, the choice of $(m,n)$ is unique; however, when $D$ increases, we can have different sets of $(m,n)$ for each $D$.  For example in $D=8$, there are six $x^i$'s, and hence we can either use two $SU(2)$ triplets with each occupying 3 of the six dimensions, or use three $SU(2)$ triplets, but with only two gauge fields turned for each triplet. We enumerate all the possible combinations of $(m,n)$ and the number of Yang-Mills fields needed up to $D=11$ dimension in the following table.
\be
\begin{tabular}{|c|c|c|c|c|c|c|c|c|c|c|c|}
  \hline
  $D$ & 4 & 5 & 6 & 7 & \multicolumn{2}{|c|}{8} & 9 & \multicolumn{2}{|c|}{10} & \multicolumn{2}{|c|}{11}\\
  \hline
  $(m,n)$ & (1,0) & (0,1) & (2,0) & (1,1) & (0,2) & (3,0) & (2,1) & (1,2) & (4,0) & (3,1) & (0,3)\nn\\
  \hline
  $N$ & 1 & 1 & 2 & 2 & 2 & 3 & 3 & 3 & 4 & 4 & 3\\
  \hline
\end{tabular}
\ee

Although the Maxwell field cannot support the Lifshitz vacua, they can be used to construct charged black holes that are asymptotic to the Lifshitz vacua. The ansatz for the Maxwell field is $\mathcal{A}=\varphi(r)dt$. Then the Maxwell equation implies that
\be
\varphi'=\frac{q}{r^{D-2}}\sqrt{\frac{h}{f}}.
\ee
In this paper, a prime denotes a derivative with respect to $r$.

Substituting all of the ansatz into the equations of motion (\ref{eom21}) and (\ref{eom22}), we find that the full set of equations are now reduced to three independent ones:
\be
\psi''+\left(\frac{h'}{2h}+\frac{f'}{2f}+\frac{D-4}{r}\right)\psi'-\frac{2\psi^3}{r^2f}=0,
\ee
\be
\frac{f'}{f}-\frac{h'}{h}+\frac{2\psi'^2}{g^2_sr}=0,
\ee
\be
\frac{f'}{f}+\frac{h'}{h}+\frac{2\psi^4}{g^2_sr^3f}+\frac{2q^2}{(D-2)r^{2D-5}f}+\frac{4\Lambda r}{(D-2)f}+\frac{2(D-3)}{r}=0.
\ee
Interestingly, these equations depend only on the total dimension $D$, rather than the individual $(m,n)$.  The equations for $D=4,5$ were obtained in \cite{Fan:2014ixa}.

\subsection{Colored Lifshitz vacua}

We first turn off the Maxwell field by setting $q=0$. It is straightforward to show that there exist a class of Lifshitz spacetimes with dynamical critical exponent $z>1$:
\be
h=\ell^2r^{2z},\quad f=\frac{r^2}{\ell^2},\quad \psi=\sqrt{\frac{z+D-3}{2\ell^2}}r,\quad\varphi=0,
\ee
with
\be
\Lambda=-\frac{(D-2)(z^2+(D-2)z+(D-1))}{4\ell^2},\quad g^2_s=\frac{z+D-3}{2(z-1)\ell^2}.
\ee
After performing a scaling $x^i_I\rightarrow\ell x^i_I$, the solution takes the form
\bea
ds^2&=&\ell^2\left(-r^{2z}dt^2+\frac{dr^2}{r^2}+r^2dx^i_Idx^i_I\right),\nn\\
\psi&=&pr,\quad p=\sqrt{\frac{z+D-3}{2}}.
\eea
Note that the condition for $z>1$ is required by the positiveness of the Yang-Mills coupling constant square $g_s^2>0$. The solutions for $D=4,5$ were obtained in \cite{Fan:2014ixa}.

\subsection{Charged Lifshitz black holes}

Having obtained the colored Lifshitz vacua in Einstein-Yang-Mills theory in general dimensions, we would like to construct black holes that are asymptotic to these vacua. This can be done by turning on the Maxwell field. We obtain a class of exact solutions for $z=D-1$, given by
\be
ds^2=\ell^2\left(-r^{2(D-1)}\tilde{f}dt^2+\frac{dr^2}{r^2\tilde{f}}+r^2dx^i_Idx^i_I\right),\quad \tilde{f}=1-\frac{q^2\ell^2}{(D-2)r^{2(D-2)}},
\ee
with
\be
g^2_s=\frac{1}{\ell^2},\quad \Lambda=-\frac{(D-1)^2(D-2)}{2\ell^2}.
\ee
The gauge potentials of Yang-Mills and Maxwell fields are given by
\be
\psi=\sqrt{D-2}r,\quad \varphi=\varphi_0+\ell^2qr.
\ee
Here $\varphi_0$ is a constant which can be viewed as the gauge parameter or simply as an integration constant.  We shall choose a gauge such that ${\cal A}$ vanishes on the horizon.  The solutions describe black holes with event horizons at $r=r_0>0$ where $\tilde f(r_0)=0$.

In the following, we set $\ell=1$ for simplicity without loss of generality. The temperature and entropy can be calculated using the standard method, given by
\be
T=\frac{1}{2\pi}(D-2)r^{D-1}_0,\quad S=\ft14\omega r_0^{D-2},\label{specialts}
\ee
where $\omega=\int dx^i_I dx^i_I$.  The electric charge can be easily obtained,
\be
Q=\frac{1}{8\pi}\int_{r\rightarrow\infty}\ast\mathcal{F}=\frac{\omega}{8\pi}q.
\ee
There is a subtlety in the definition of electric potential. Following the discussion in \cite{Fan:2014ixa,Liu:2014dva}, we define directly
\be
\Phi=\varphi_0=-qr_0
\ee
to be the electric potential. (Note that we have chosen a gauge such that $\mathcal{A}$ vanishes on the horizon.) It is easy to check that
\be
TdS+\Phi dQ=0.\label{fl}
\ee
This is the thermodynamical first law for our black holes constructed above. We find that the mass parameter of these black holes is zero, so these black holes are charactered only by the electric charge $Q$.

\section{Hyperscaling violating Lifshitz black holes with a dilaton}

The concept of hypercaling violation has been developed in condensed matter and in the context of gauge gravity duality. It provides a new holographic realization. In the following we can see, by including dilaton, the theory supports a class of asymptotic Lifshitz spacetimes with an overall hyperscaling violating factor.

\subsection{Colored hyperscaling violating Lifshitz vacua}

In this section, we introduce a dilaton $\phi$ to the system and propose the following Lagrangian
\be
\mathcal{L}_D=\sqrt{-g}\left(R-V(\phi)-\frac{1}{2}(\partial\phi)^2-\sum^N_{I=1}\frac{1}{4g^2_I}e^{\lambda\phi}F^2_I-\frac{1}{4}e^{\lambda\phi}\mathcal{F}^2\right).
\ee
The full set of covariant equations of motion are given by
\be
\nabla_\mu(e^{\lambda\phi}F^{a\mu\nu})+\epsilon^{abc}A^b_{\mu}e^{\lambda\phi}F^{c\mu\nu}=0,\quad\nabla_\mu(e^{\lambda\phi}\mathcal{F}^{\mu\nu})=0,\label{eom31}
\ee
\be
\Box\phi=\sum^N_{I=1}\frac{\lambda}{4g^2_I}e^{\lambda\phi}F^2_I+\frac{\lambda}{4}e^{\lambda\phi}\mathcal{F}^2+\frac{\partial V}{\partial\phi},\label{eom32}
\ee
\begin{align}
R_{\mu\nu}=&\frac{1}{2}\partial_\mu\phi\partial_\nu\phi+\sum^N_{I=1}\frac{1}{2g^2_I}e^{\lambda\phi}\left(g^{\rho\sigma}F^a_{I\mu\rho}F^a_{I\nu\sigma}-\frac{1}{2(D-2)}F^2_Ig_{\mu\nu}\right)\nn\\
&+\frac{1}{2}e^{\lambda\phi}\left(g^{\rho\sigma}\mathcal{F}_{\mu\rho}\mathcal{F}_{\nu\sigma}-\frac{1}{2(D-2)}\mathcal{F}^2g_{\mu\nu}\right)+\frac{V}{D-2}g_{\mu\nu}.\label{eom33}
\end{align}
Motivated by string theory, we consider the following exponential behavior of the scalar potential,
\be
V=\Lambda e^{-\lambda\phi},
\ee
and ansatz for the scalar,
\be
\phi=\frac{\theta}{\lambda}\log r,
\ee
where $\theta$ is the hyperscaling parameter. Meanwhile, the ansatz for Maxwell field is $\mathcal{A}=\varphi dt$. Then the Maxwell equation implies
\be
\varphi'=\frac{q}{r^{\frac{(D-2)(\theta+2)}{2}}}\sqrt{\frac{h}{f}}.
\ee
Now we consider a class of solutions whose metrics take the form
\be
ds^2=r^\theta\left(-h(r)dt^2+\frac{dr^2}{f(r)}+r^2dx^i_Id^i_I\right).
\ee
Using the same ansatz for Yang-Mills fields, the equations of motion (\ref{eom31}), (\ref{eom32}), (\ref{eom33}) now reduce to
\be
\psi''+\left(\frac{f'}{2f}+\frac{h'}{2h}+\frac{(D-2)\theta}{2r}+\frac{D-4}{r}\right)\psi'-\frac{2\psi^3}{r^2f}=0,
\ee
\be
\left(1+\frac{\theta}{2}\right)\left(\frac{f'}{f}-\frac{h'}{h}\right)+\frac{\psi'^2}{g^2_sr}+\frac{\theta^2}{(D-2)\lambda^2r}-\frac{\theta^2}{2r}-\frac{\theta}{r}=0,
\ee
\begin{align}
\left(1+\frac{\theta}{2}\right)\left(\frac{f'}{f}+\frac{h'}{h}\right)&+\frac{\psi^4}{g^2_sr^3f}+\frac{q^2}{(D-2)r^{(D-2)(\theta+2)-1}f}+\frac{2\Lambda r}{(D-2)f}\nn\\
&+\frac{(D-2)\theta^2}{2r}+\frac{(2D-5)\theta}{r}+\frac{2(D-3)}{r}=0.
\end{align}
We find the following Lifshitz vacua with hyperscaling violation
\bea
ds^2&=&r^\theta\left(-r^{2z}dt^2+\frac{dr^2}{r^2}+r^2dx^i_Idx^i_I\right),\nn\\
\psi&=&\sqrt{\frac{2z+(D-2)\theta+2(D-3)}{4}}r,\quad\varphi=0,\nn\\
\lambda&=&\sqrt{\frac{2\theta}{(D-2)(2z+\theta)}},\quad g^2_s=\frac{2z+(D-2)\theta+2(D-3)}{8(z-1)},\nn\\
\Lambda&=&-\frac{D-2}{4}\{(D-2)\theta^2+[Dz+3(D-2)]\theta+2[z^2+(D-2)z+(D-1)]\},
\eea
when $\theta=0$, the dilaton decouples from the theory, so it reduces to the Einstein-Yang-Mills-Maxwell theory. The solution reduces to the one obtained in section 2.2.

\subsection{Charged hyperscaling violating Lifshitz black holes}

Naturally, it is time to construct black holes solutions. Also this can be done by turning on the Maxwell field. We find that when
\be
\theta=\frac{2}{D-2}[z-(D-1)],
\ee
we can obtain a class of charged Lifshitz black holes solutions with hyperscaling violation, given by
\be
ds^2=r^\theta\left(-r^{2z}\tilde{f}dt^2+\frac{dr^2}{r^2\tilde{f}}+r^2dx^i_Idx^i_I\right),\quad\tilde{f}=1-\frac{q^2}{2(z-1)r^{2(z-1)}},
\ee
with
\be
\lambda=\sqrt{\frac{2[(z-(D-1)]}{(D-1)(D-2)(z-1)}},\quad g_s=\frac{1}{2},\quad \Lambda=-[D(z-1)^2+z-1].
\ee
The various fields are given by
\be
\psi=\sqrt{z-1}r,\quad\varphi=\varphi_0+qr,\quad\phi=\sqrt{\frac{2(D-1)}{D-2}(z-1)[z-(D-1)]}\log r,
\ee
when $\theta=0$, corresponding to $z=D-1$, also the dilaton decouples. The solution reduces to the one obtained in section 2.3.

These black holes satisfy the same thermodynamical first law (\ref{fl}), with
\be
T=\frac{z-1}{2\pi}r^z_0,\quad S=\frac{1}{4}\omega r^{z-1}_0,\quad Q=\frac{\omega}{16\pi}q,\quad \Phi=-qr_0,
\ee
where $r_0$ is the horizon, satisfying $\tilde{f}(r_0)=0$.
\section{Conclusions}

In this paper, we generalized the work in \cite{Fan:2014ixa,Fan:hv} and constructed non-abelian Lifshitz spacetimes with dynamic exponents $z>1$ in general dimensional Einstein gravities coupled to a cosmological constant and $N$ $SU(2)$ Yang-Mills fields. We also introduced a Maxwell field and constructed exact charged black holes when $z=D-1$. Furthermore, we introduced a dilaton to the system and construct Lifshitz spacetimes with hyperscaling violations. Analogously, we obtained exact charged black holes with hyperscaling violation when $\theta=\frac{2}{D-2}[z-(D-1)]$ after turning on the Maxwell field. All the black holes we constructed satisfy a non-trival thermodynamical first law (\ref{fl}) with vanishing mass, both the Yang-Mills fields and dilaton did not modify it.

The $N$ $SU(2)$ Yang-Mills fields can be viewed as subgroups of a sufficiently large group.  It establishes the principle that non-abelian Lifshitz black holes can exist in arbitrary higher dimensions provided that the Yang-Mills group is sufficiently large.  It is of great interest to investigate the classification of such Lifshitz spacetimes for general Lie groups.
Furthermore, the Lifshitz spacetimes we constructed provide abundant background for studying strongly coupled system.

\section*{Acknowledgement}

We are grateful to Hong L\"u for useful assistance. This work is supported in part by NSFC grants NO.11175269, NO.11475024 and NO.11235003.

\end{document}